\documentclass[a4paper,10pt,conference]{IEEEtran}
\IEEEoverridecommandlockouts
\usepackage{cite}
\usepackage{amsmath,amssymb,amsfonts}
\usepackage{graphicx}
\usepackage{textcomp}
\usepackage{xcolor}
\usepackage[caption=false,font=footnotesize]{subfig}
\def\BibTeX{{\rm B\kern-.05em{\sc i\kern-.025em b}\kern-.08em
    T\kern-.1667em\lower.7ex\hbox{E}\kern-.125emX}}
\begin{document}

\title{Security Evaluation of Block-based Image Encryption for Vision Transformer against Jigsaw Puzzle Solver Attack}

\author{\IEEEauthorblockN{1\textsuperscript{st} Tatsuya Chuman}
\IEEEauthorblockA{\textit{Tokyo Metropolitan University} \\
Tokyo, Japan \\
chuman-tatsuya1@ed.tmu.ac.jp}
\and
\IEEEauthorblockN{2\textsuperscript{nd} Hitoshi Kiya}
\IEEEauthorblockA{\textit{Tokyo Metropolitan University} \\
Tokyo, Japan \\
kiya@tmu.ac.jp}
}
\makeatletter
\def\ps@IEEEtitlepagestyle{%
  \def\@oddfoot{\mycopyrightnotice}%
  \def\@evenfoot{}%
}
\def\mycopyrightnotice{%
  {\footnotesize Preprint: IEEE 4th Global Conference on Life Sciences and Technologies (LifeTech 2022)\hfill}
  \gdef\mycopyrightnotice{}
}
\maketitle

\begin{abstract}
The aim of this paper is to evaluate the security of a block-based image encryption for the vision transformer against jigsaw puzzle solver attacks.
The vision transformer, a model for image classification based on the transformer architecture, is carried out by dividing an image into a grid of square patches. Some encryption schemes for the vision transformer have been proposed by applying block-based image encryption such as block scrambling and rotating to patches of the image.
On the other hand, the security of encryption scheme for the vision transformer has never evaluated.
In this paper, jigsaw puzzle solver attacks are utilized to evaluate the security of encrypted images by regarding the divided patches as pieces of a jigsaw puzzle. In experiments, an image is resized and divided into patches to apply block scrambling-based image encryption, and then the security of encrypted images for the vision transformer against jigsaw puzzle solver attacks is evaluated.
\end{abstract}


\section{Introduction}
The spread use of deep neural networks (DNNs) has greatly contributed to solving complex tasks for many applications\cite{lecun2015deeplearning}, including privacy-sensitive security-critical ones such as facial recognition and medical image analysis.
Various perceptual encryption methods have been proposed to generate visually-protected images\cite{Warit_IEEEtrans}\cite{AprilPyone_IEEEtrans}.
Although information theory-based encryption (like RSA and AES) generates a ciphertext, images encrypted by the perceptual encryption methods can be directly applied to some image processing algorithms or image compression algorithms\cite{kiya2022overview}.
Numerous encryption schemes have been proposed for privacy-preserving DNNs, but several attacks including DNN-based ones were shown to restore visual information from encrypted images\cite{ito_IEEEaccess}\cite{Jialong_ICCC}.
Therefore, encryption schemes that are robust against various attacks are essential for privacy-preserving DNNs.
\par
Although block scrambling is well-known to enhance robustness against attacks\cite{Chuman_IEEEtrans}, but it decreases the performance of DNN modes.
The use of the vision transformer, a model for image classification based on the transformer architecture, enables image encryption to apply block scrambling without a decrease in the performance of DNNs.
However, jigsaw puzzle solver attacks are considered by regarding the blocks of an encrypted image as pieces of a jigsaw puzzle\cite{CHUMAN_IEICE}\cite{Chuman_IEEEtrans}.
Accordingly, in this paper, the security of block-based image encryption for the vision transformer against the jigsaw puzzle solver attacks is evaluated.

\begin{figure}[t]
	\centering
	\includegraphics[width = \linewidth]{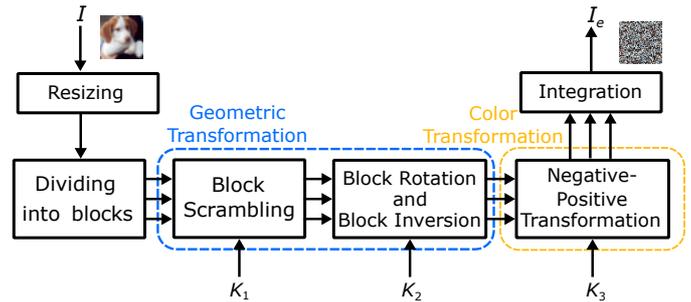}
	\caption{Block-based image encryption for vision transformer}
	\label{fig:vit_flow}
	\end{figure}

\begin{figure*}[t]
	\centering
	\captionsetup[subfigure]{justification=centering}
	\subfloat[CIFAR-10]{
	\includegraphics[width =17.5cm]{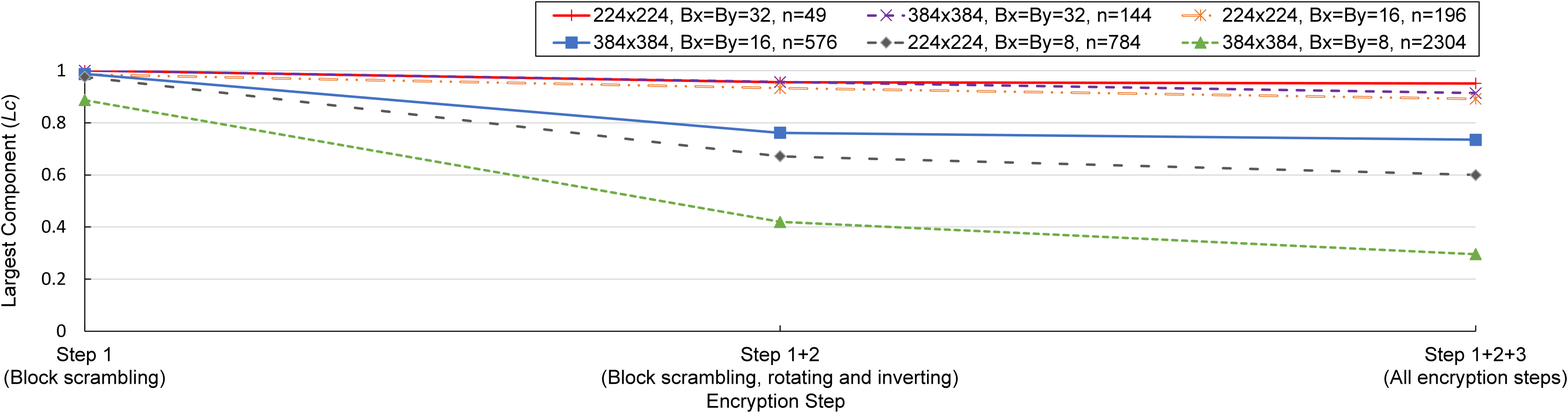}
	\label{fig:cifar10_lc}
	}
	\newline
	\subfloat[ImageNet]{
	\includegraphics[width =17.5cm]{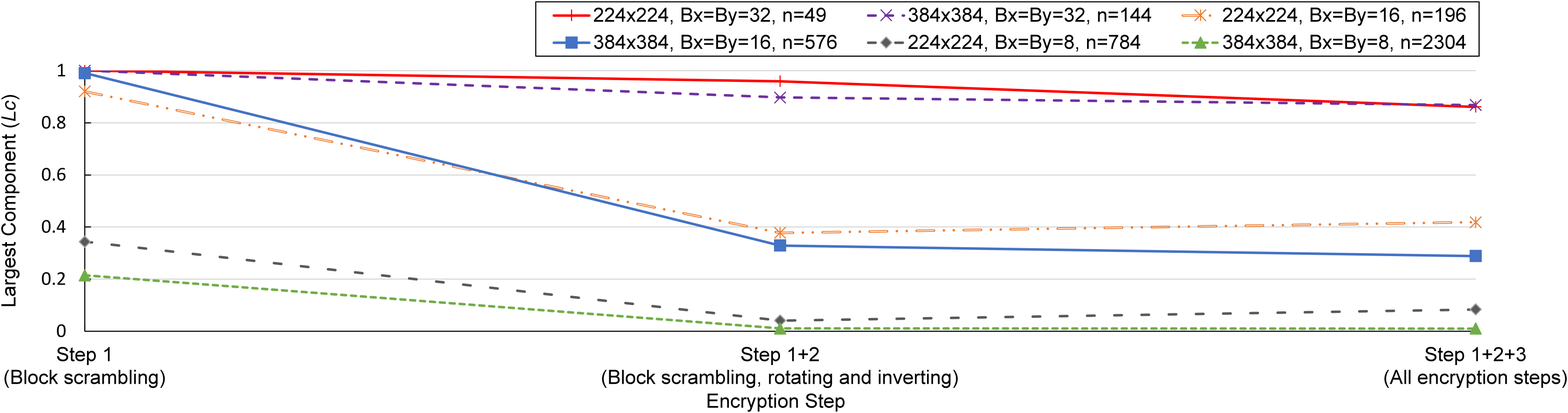}
	\label{fig:imagenet_lc}
	}
	\newline
	\caption{Security evaluation of block-based image encryption against jigsaw puzzle solver attack}
	\label{fig:result_lc}
\end{figure*}

\section{Preparation}
\subsection{Block-based Image Encryption for Vision Transformer}
It has been known that the vision transformer, a model for image classification based on the transformer architecture, is carried out by dividing an image into a grid of square patches\cite{dosovitskiy2021an}.
For example, images from the CIFAR-10 dataset are resized from 32$\times$32 to 224$\times$224 or 384$\times$384, and then divided into 16$\times$16 patch to fit the same patch size of pre-trained model such as ViT-B/16 and ViT-L/16.
\par
On the other hand, some encryption schemes for the vision transformer have been proposed by applying block-based image encryption such as block scrambling and rotating to patches of the image.
As illustrated in Fig. \ref{fig:vit_flow}, the procedure for performing image encryption to generate an encrypted image $I_e$ from an original image $I$ is given as follows.
\begin{itemize}
\setlength{\leftskip}{0.5cm}
\item[Step 1:] Divide an image with $X \times Y$ pixels into blocks, each with $B_x \times B_y$ pixels, and permute randomly the divided blocks using a random integer generated by a secret key $K_1$, where $K_1$ is commonly used for all color components.
Hence, the number of blocks $n$ is given by
\begin{equation}
n = \lfloor \frac{X}{B_x} \rfloor \times \lfloor \frac{Y}{B_y} \rfloor
\end{equation}
where $\lfloor \cdot \rfloor$ is the function that rounds down to the nearest integer.
In this paper, $B_{x}=B_{y}=8, 16,$ and $32$ are used to fit the patch size of pre-trained model\cite{dosovitskiy2021an}.
\item[Step 2:] Rotate and invert randomly each block by using a random integer generated by a key $K_2$, where $K_2$ is commonly used for all color components as well.
\item[Step 3:] Apply negative-positive transformation to each block by using a random binary integer generated by a key $K_3$, where $K_3$ is commonly used for all color components. In this step, a transformed pixel value in the $i$th block $B_i$, $p'$, is computed using
\begin{equation}
p'=
\left\{
\begin{array}{ll}
p & (r(i)=0) \\
p \oplus (2^L-1) & (r(i)=1)
\end{array} ,
\right.
\end{equation}
where $r(i)$ is a random binary integer generated by $K_3$, and $p \in B_i$ is
the pixel value of the original image with $L$ bit per pixel.
In this paper, the value of occurrence probability $P(r(i))=0.5$ has been used to invert bits randomly.
\end{itemize}

\subsection{Jigsaw Puzzle Solver Attacks}
Some jigsaw puzzle solver attacks have been proposed to assemble images encrypted with the block-based image encryption\cite{CHUMAN2017ICME}.
It has been known that robustness against jigsaw puzzle solver attacks is enhanced when encrypted images have a large number of blocks and a block size of ones is small.
On the other hand, images encrypted with the block-based scheme for vision transformer have never evaluated.
Thus, in this paper, we evaluate the security of block-based image encryption for vision transformer against the jigsaw puzzle solver.

\begin{figure*}[t]
	\captionsetup[subfigure]{justification=centering}
	\centering
	\begin{center}
	\subfloat[Original image]
	{
	\includegraphics[height=3.1cm]{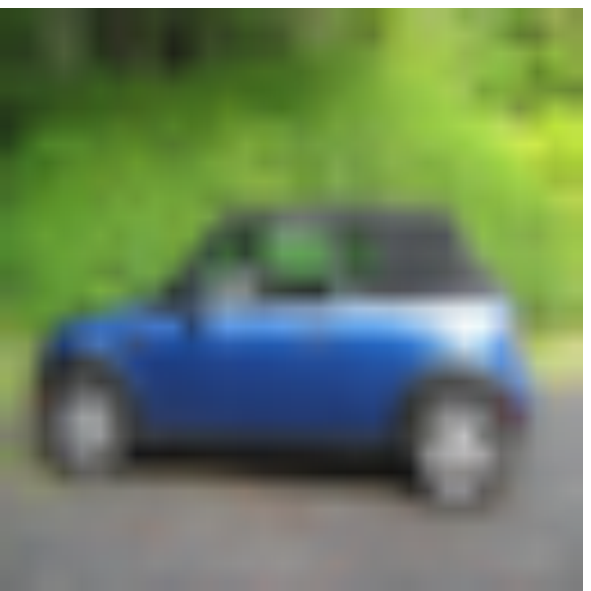}
	\label{fig:car_orig}
	}
	\subfloat[$X\!\times\!Y\!=\!224\!\!\times\!\!224$,\\$B_{x}=B_{y}=32$,\\$n=49$,\\$L_{c}=0.96$]
	{
	\includegraphics[height=3.1cm]{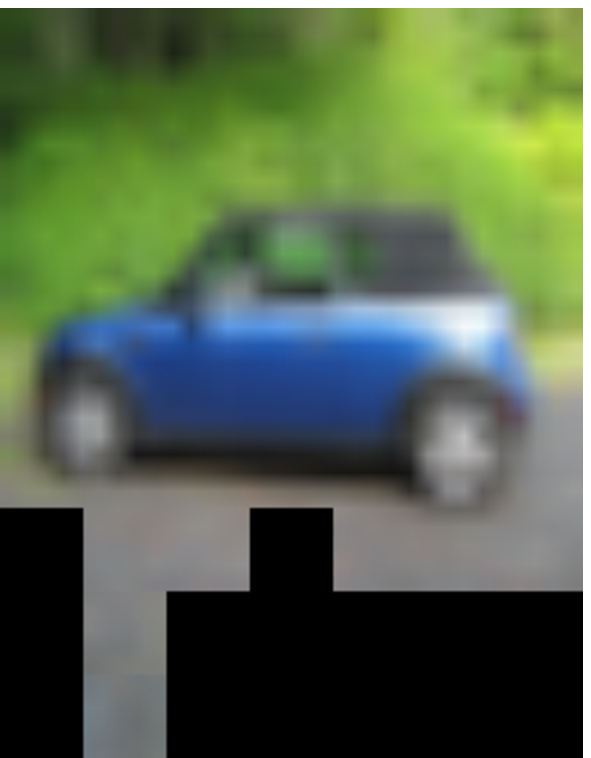}
	\label{fig:224_32_car}
	}
	\subfloat[$X\!\times\!Y\!=\!384\!\!\times\!\!384$,\\$B_{x}=B_{y}=32$,\\$n=144$,\\$L_{c}=0.97$]
	{\label{fig:384_32_car}
	\includegraphics[height=3.1cm]{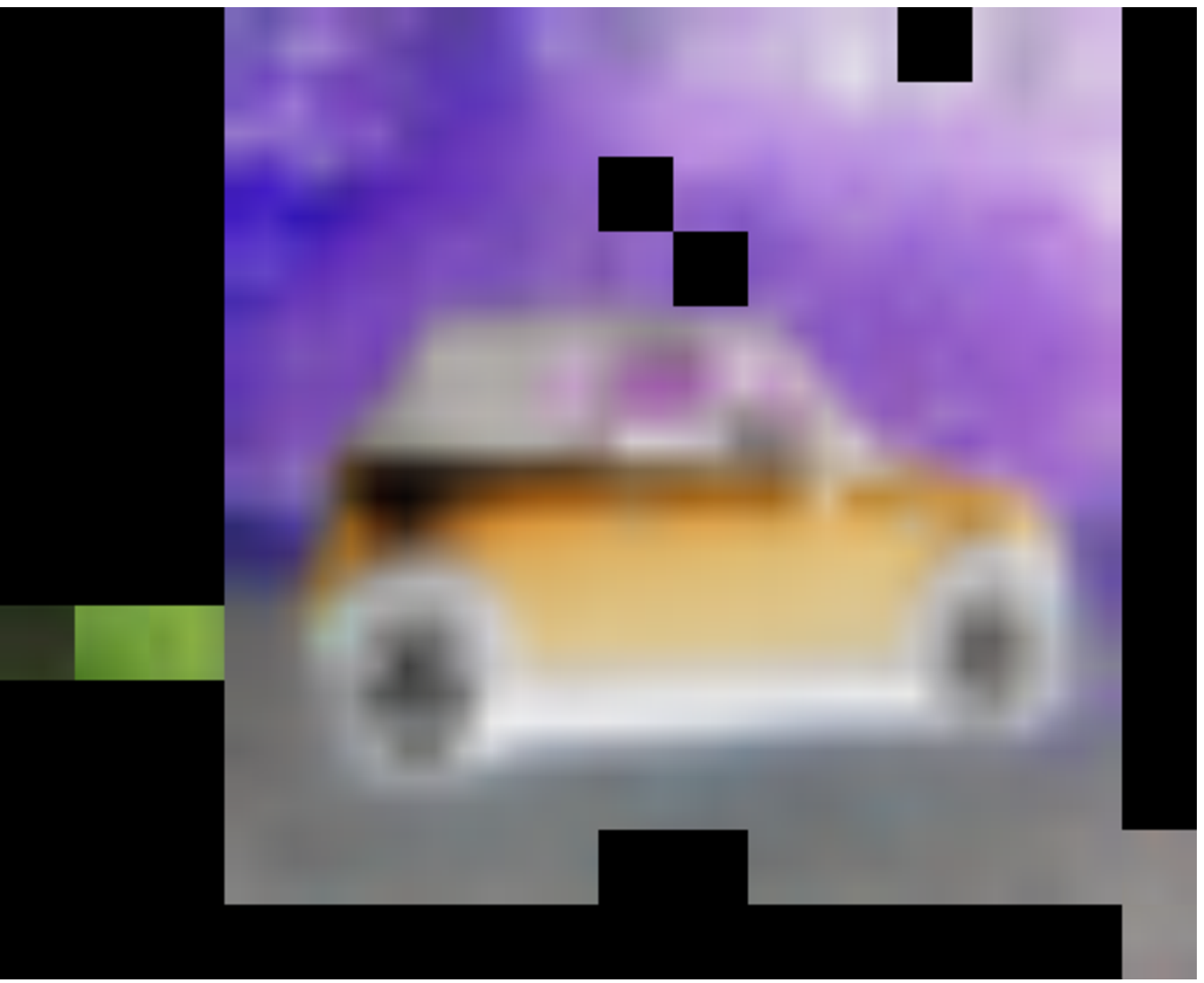}
	}	
	\subfloat[$X\!\times\!Y\!=\!224\!\!\times\!\!224$,\\$B_{x}=B_{y}=16$,\\$n=196$,\\$L_{c}=0.93$]
	{\label{fig:224_16_car}
	\includegraphics[height=3.1cm]{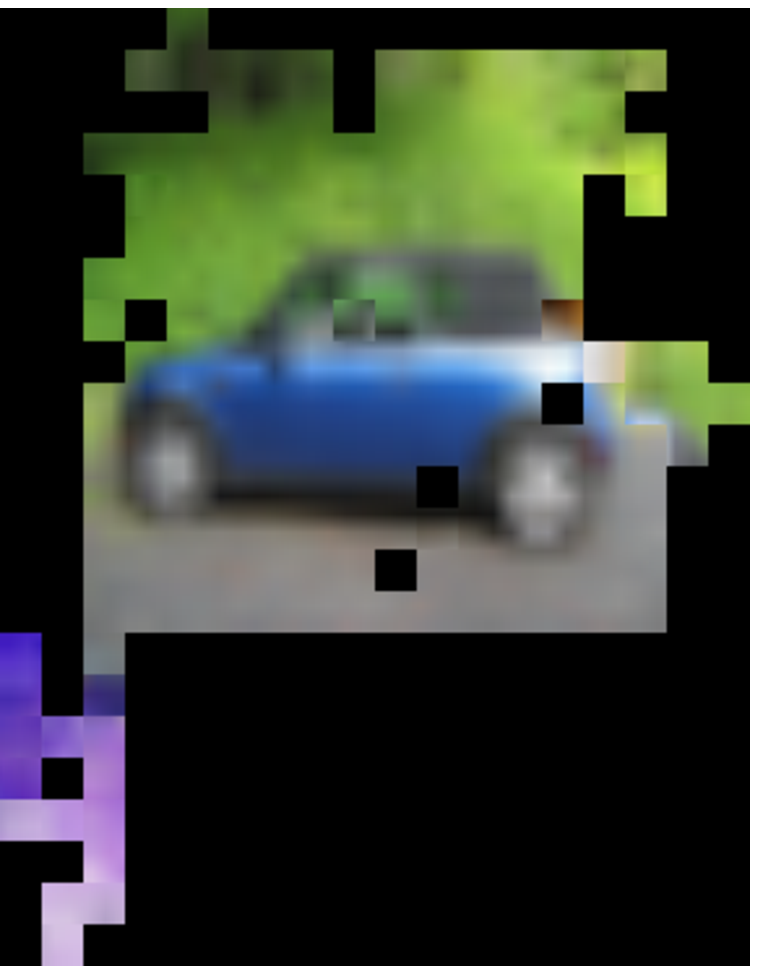}
	}
	\\
	\subfloat[$X\!\times\!Y\!=\!384\!\!\times\!\!384$,\\$B_{x}=B_{y}=16$,\\$n=576$,\\$L_{c}=0.76$]
	{\label{fig:384_16_car}
	\includegraphics[height=3.1cm]{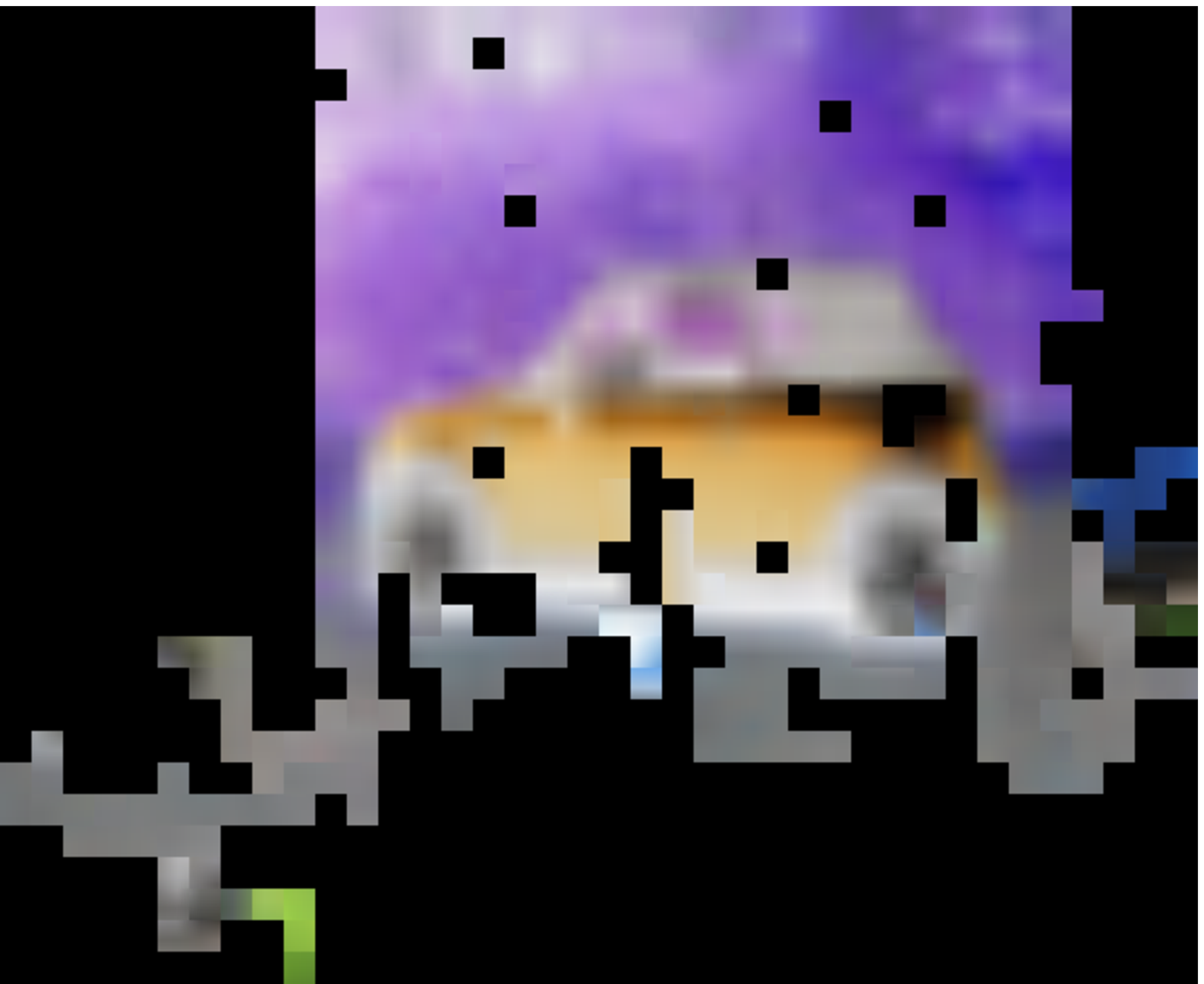}
	}
	\subfloat[$X\!\times\!Y\!=\!224\!\!\times\!\!224$,\\$B_{x}=B_{y}=8$,\\$n=784$,\\$L_{c}=0.60$]
	{\label{fig:224_8_car}
	\includegraphics[height=3.1cm]{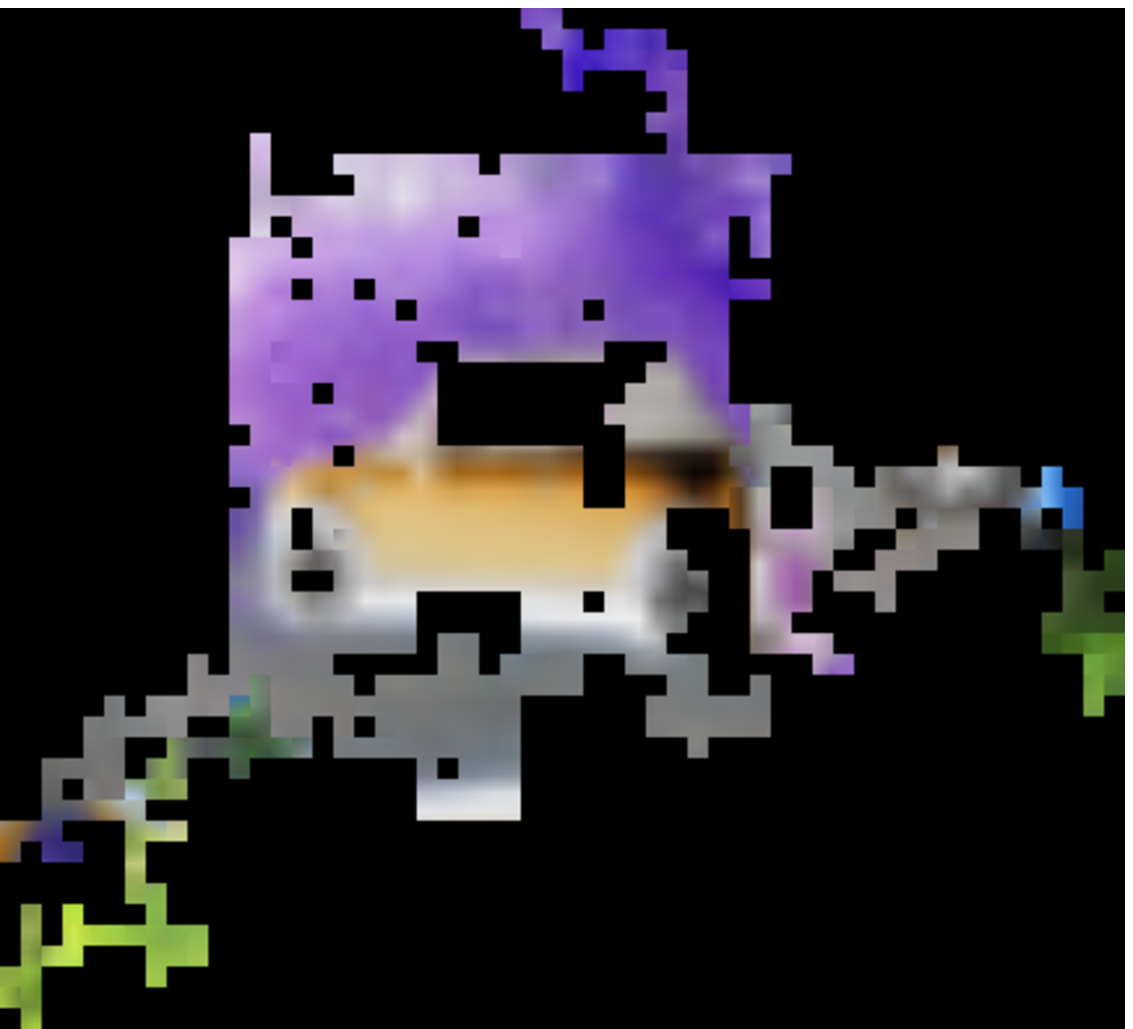}
	}
	\subfloat[$X\!\times\!Y\!=\!384\!\!\times\!\!384$,\\$B_{x}=B_{y}=8$,\\$n=2304$,\\$L_{c}=0.31$]
	{\label{fig:384_8_car}
	\includegraphics[height=3.1cm]{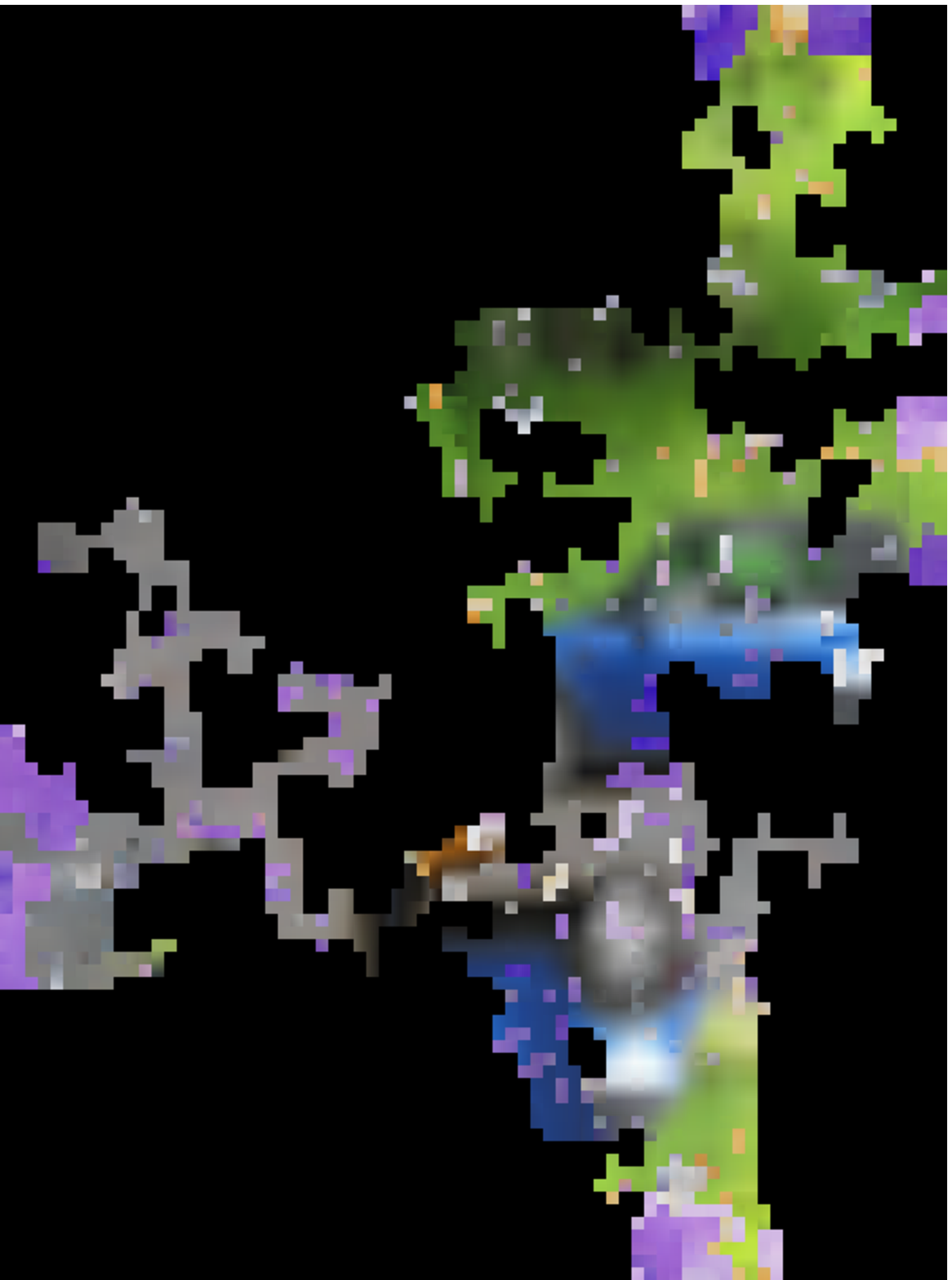}
	}
	\end{center}
	\caption{Examples of assembled images by using the jigsaw puzzle solver (CIFAR-10, Step1+2+3)}
	\label{fig:4step_car}
\end{figure*}

\begin{figure*}[t]
	\captionsetup[subfigure]{justification=centering}
	\centering
	\begin{center}
	\subfloat[Original image]
	{
	\includegraphics[height=3.1cm]{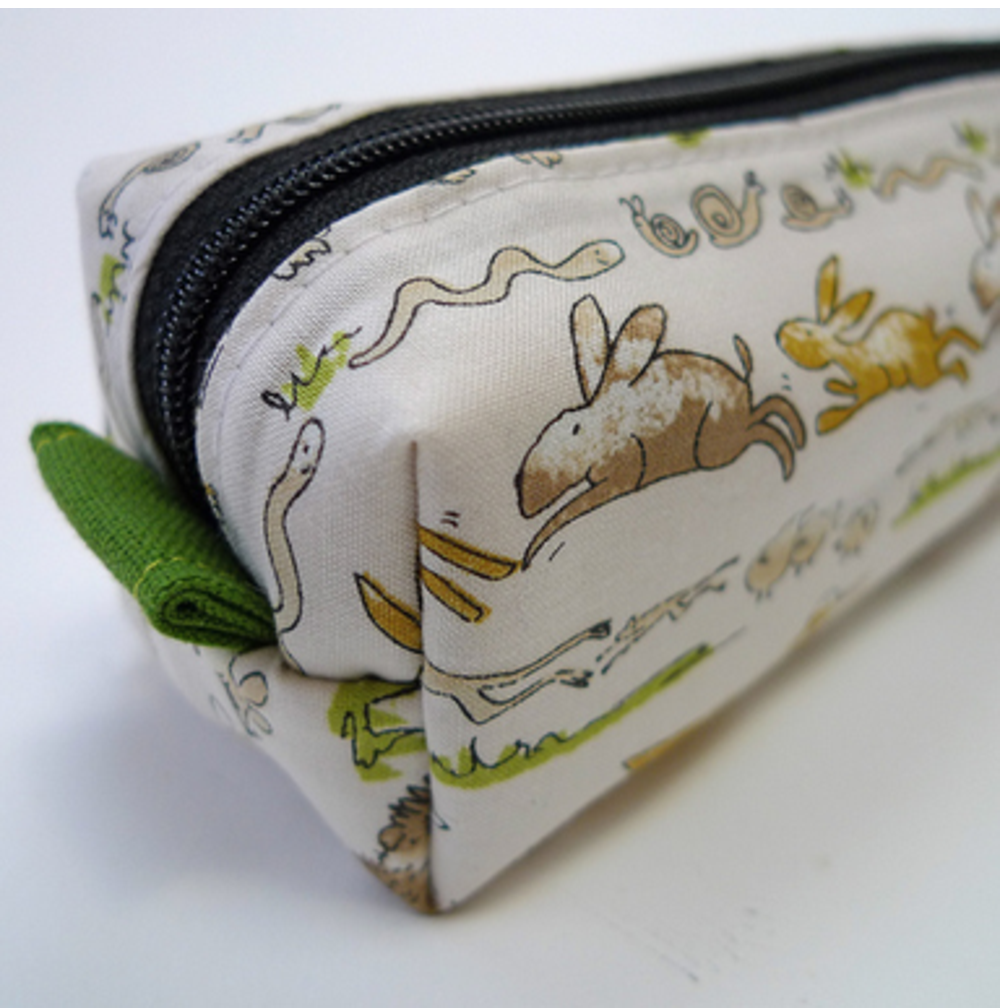}
	\label{fig:pen_orig}
	}
	\subfloat[$X\!\times\!Y\!=\!224\!\!\times\!\!224$,\\$B_{x}=B_{y}=32$,\\$n=49$,\\$L_{c}=0.61$]
	{
	\includegraphics[height=3.1cm]{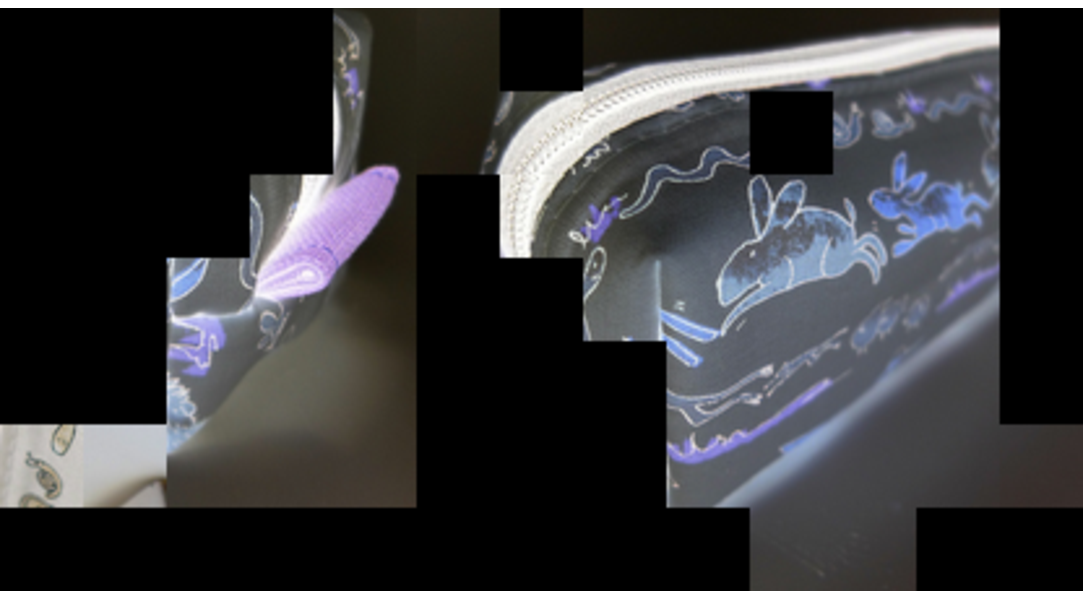}
	\label{fig:224_32_pen}
	}
	\subfloat[$X\!\times\!Y\!=\!384\!\!\times\!\!384$,\\$B_{x}=B_{y}=32$,\\$n=144$,\\$L_{c}=0.92$]
	{\label{fig:384_32_pen}
	\includegraphics[height=3.1cm]{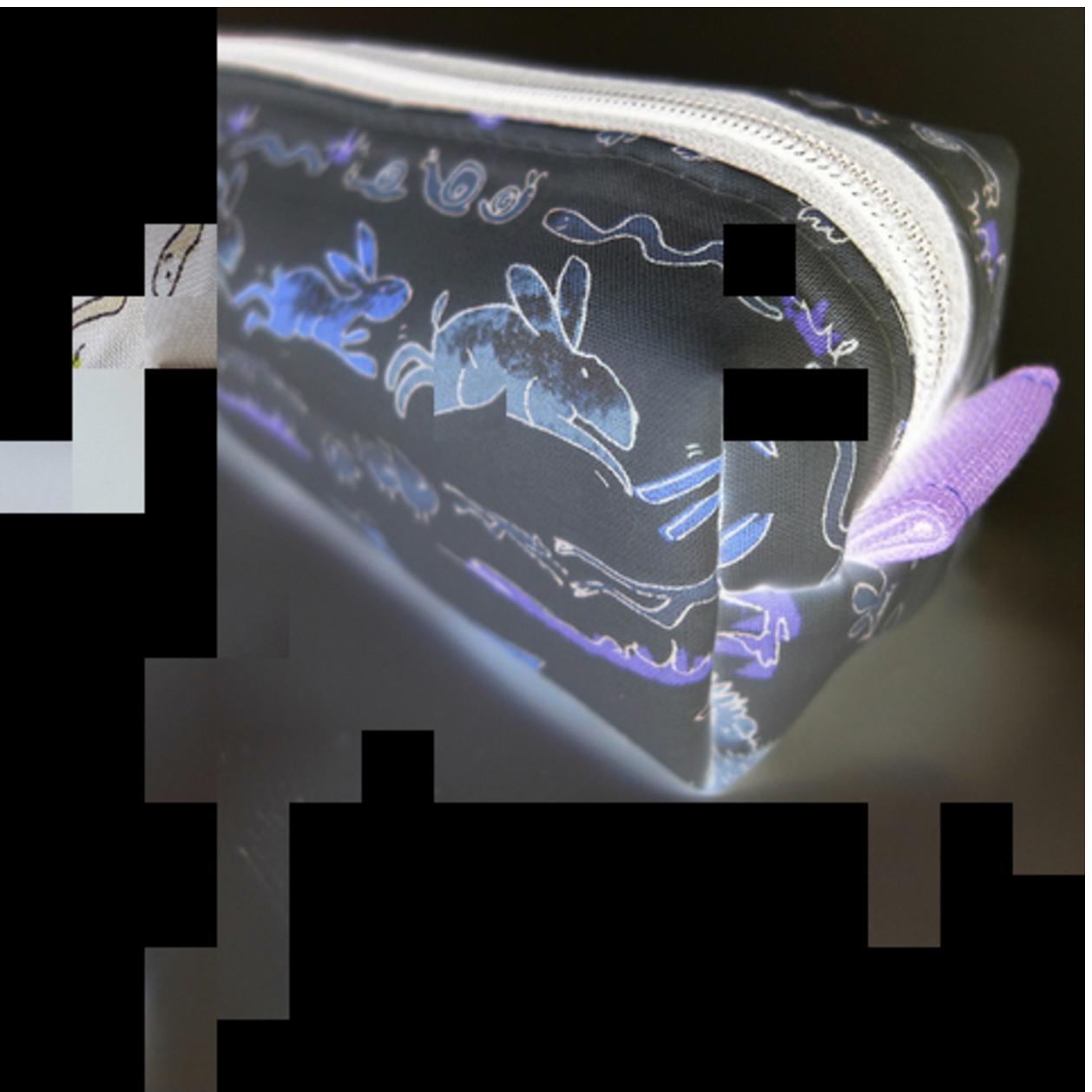}
	}	
	\subfloat[$X\!\times\!Y\!=\!224\!\!\times\!\!224$,\\$B_{x}=B_{y}=16$,\\$n=196$,\\$L_{c}=0.22$]
	{\label{fig:224_16_pen}
	\includegraphics[height=3.1cm]{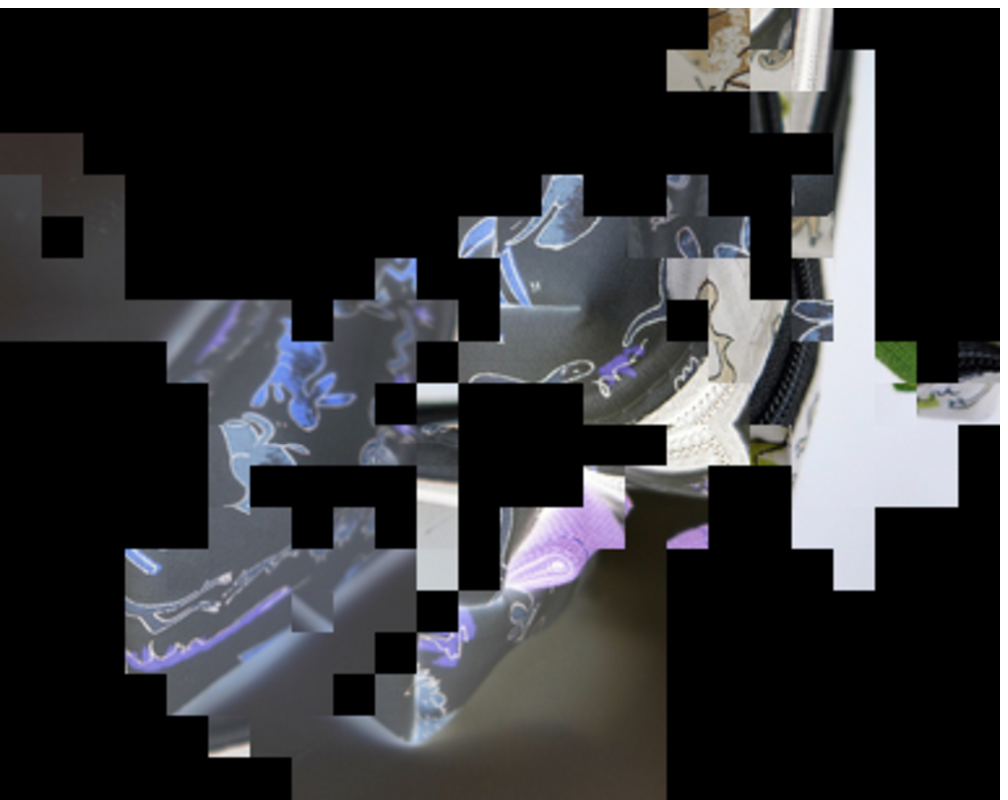}
	}
	\\
	\subfloat[$X\!\times\!Y\!=\!384\!\!\times\!\!384$,\\$B_{x}=B_{y}=16$,\\$n=576$,\\$L_{c}=0.24$]
	{\label{fig:384_16_pen}
	\includegraphics[height=3.1cm]{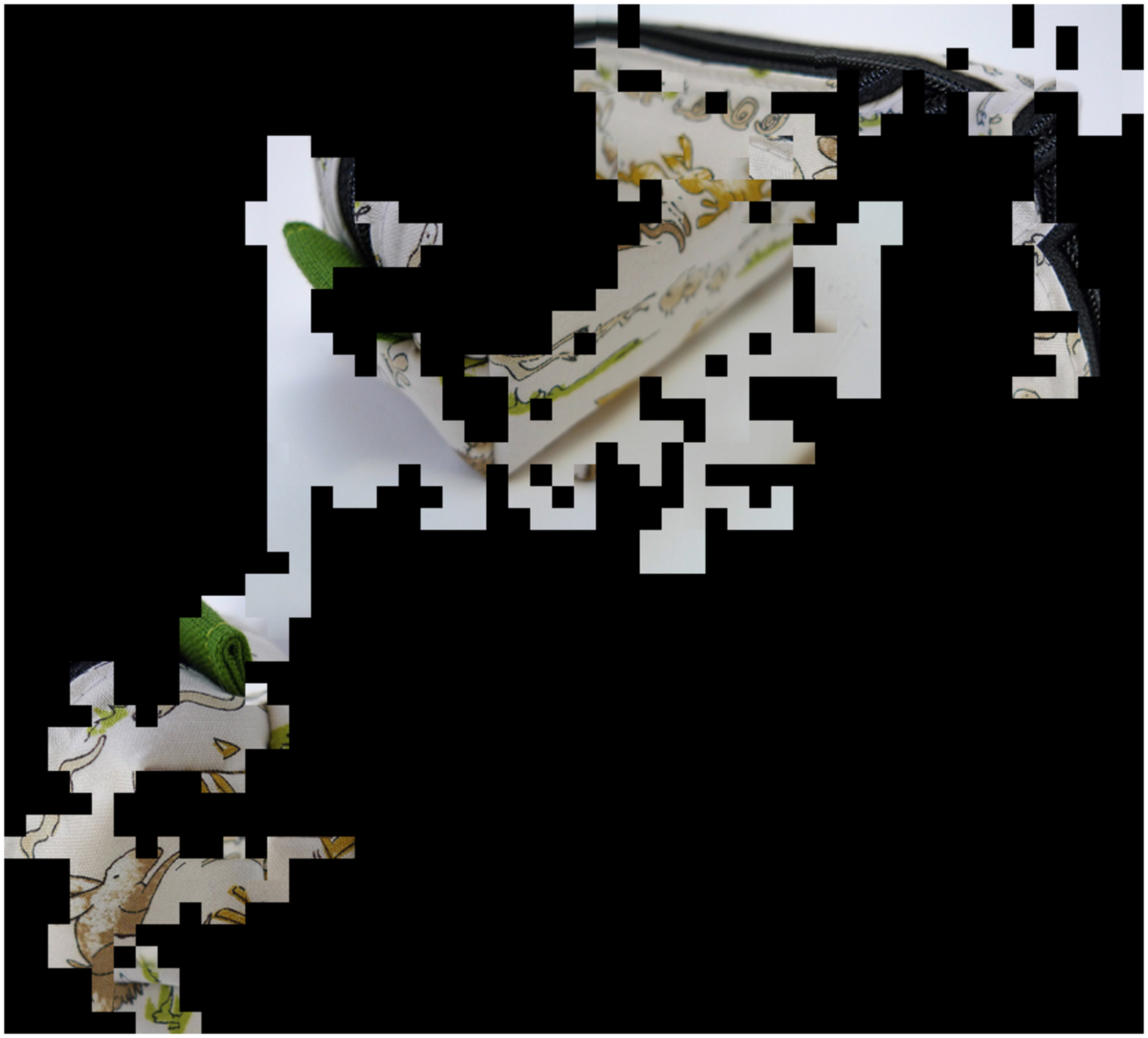}
	}
	\subfloat[$X\!\times\!Y\!=\!224\!\!\times\!\!224$,\\$B_{x}=B_{y}=8$,\\$n=784$,\\$L_{c}=0.22$]
	{\label{fig:224_8_pen}
	\includegraphics[height=3.1cm]{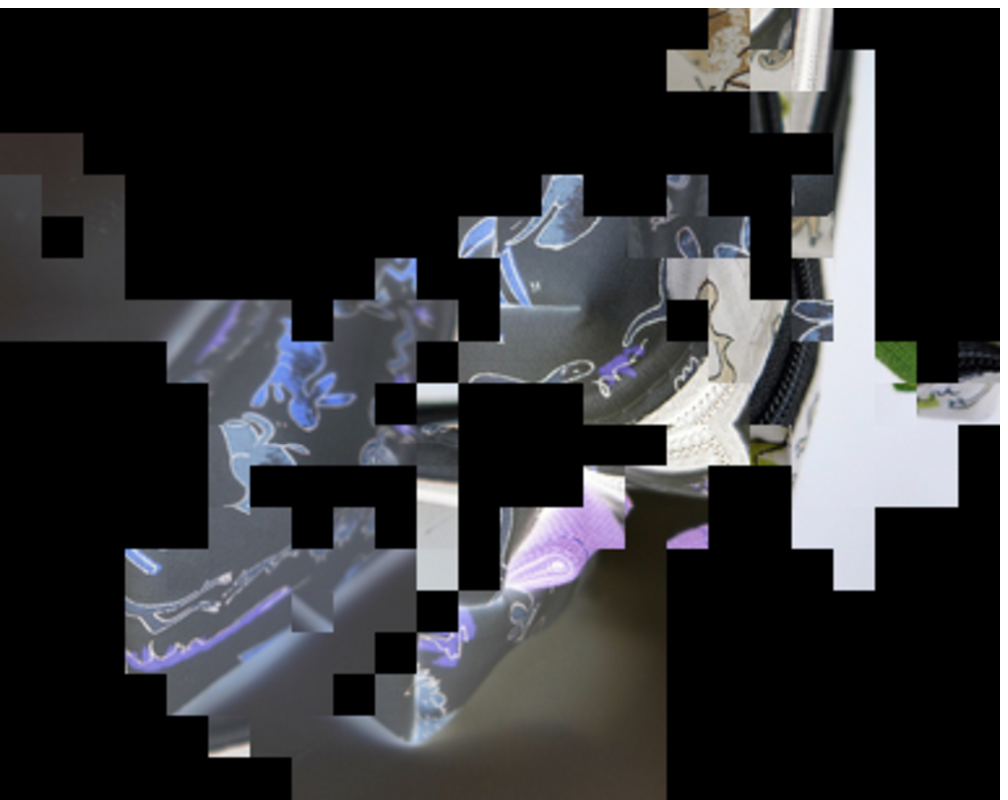}
	}
	\subfloat[$X\!\times\!Y\!=\!384\!\!\times\!\!384$,\\$B_{x}=B_{y}=8$,\\$n=2304$,\\$L_{c}=0.01$]
	{\label{fig:384_8_pen}
	\includegraphics[height=3.1cm]{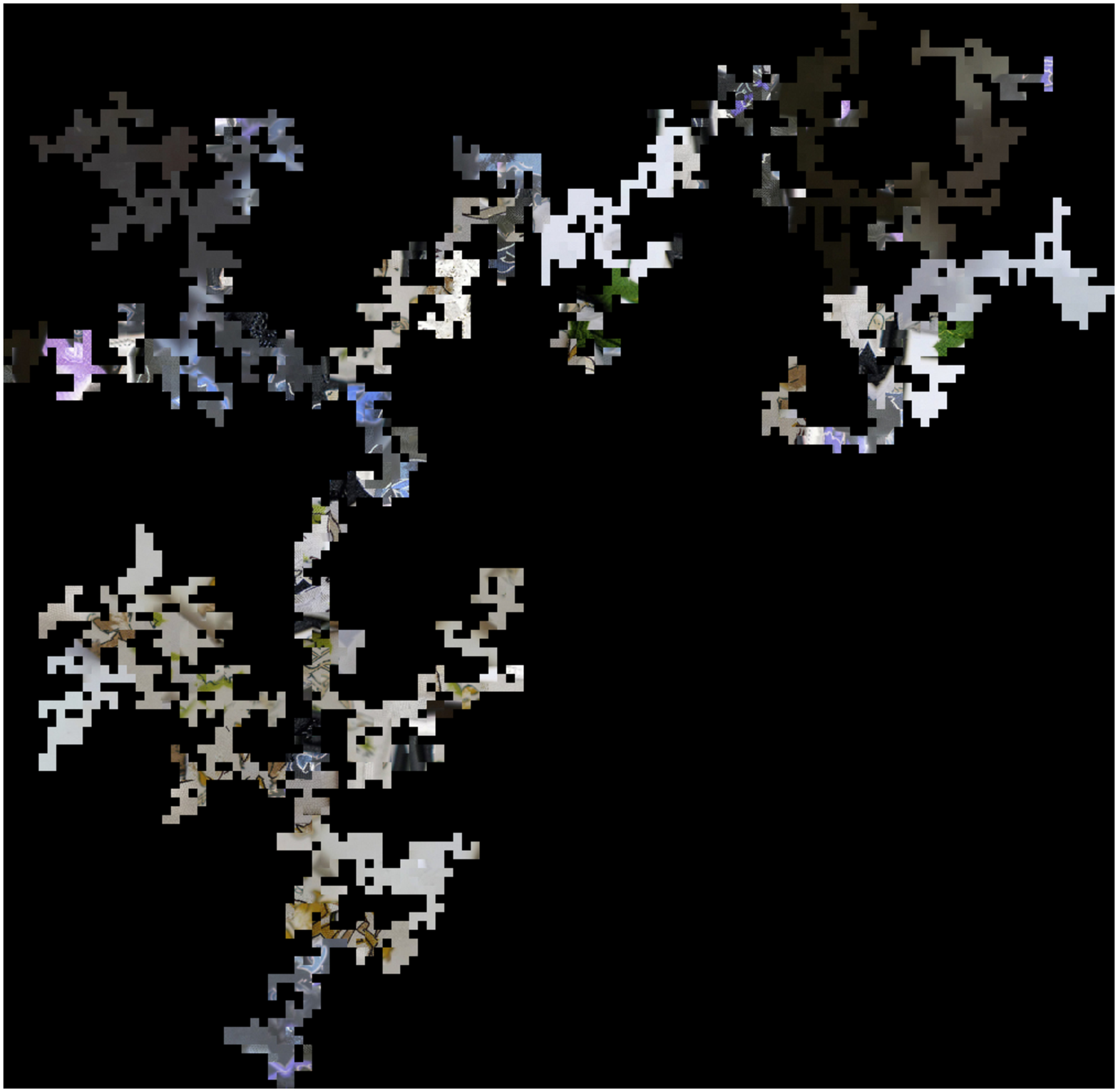}
	}
	\end{center}
	\caption{Examples of assembled images by using the jigsaw puzzle solver (ImageNet, Step1+2+3)}
	\label{fig:4step_pen}
\end{figure*}

\section{Experiments and Results}
\subsection{Experimental Conditions}
In this section, the security of block-based image encryption for the vision transformer is discussed by using the jigsaw puzzle solver\cite{CHUMAN2017ICME}.
Largest component $L_{c}$ was used to evaluate the results, which is the ratio of the number of the largest joined blocks that have correct adjacencies to the number of blocks in an image\cite{Gallagher_2012_CVPR}.
In the measure, $L_{c} \in [0,1]$, a smaller value means the difficulty of recognizing objects.
\par
We used five images randomly chosen from the CIFAR-10 and ImageNet datasets separately.
Each image from the CIFAR-10 dataset was resized from 32$\times$32 to 224$\times$224 and 384$\times$384.
On the other hand, each image from the ImageNet dataset was resized from 500$\times$500 to 224$\times$224 and 384$\times$384.
Next, five different encrypted images were generated from one ordinary image by using different keys.
We assembled the encrypted images by using the jigsaw puzzle solver and chose the image that had the highest $L_{c}$.
We performed this procedure for each encrypted image independently and calculated the average $L_{c}$ for the five images.

\subsection{Experimental Results}
Figures \ref{fig:result_lc}\subref{fig:cifar10_lc} and \subref{fig:imagenet_lc} show the security evaluation of encrypted images from the CIFAR-10 and ImageNet datasets against the jigsaw puzzle solver attack. The more encryption steps increase, the more difficult jigsaw puzzle solver assemble encrypted images.
As shown in Figs.\ref{fig:result_lc}\subref{fig:cifar10_lc} and \subref{fig:imagenet_lc}, $L_{c}$ for the encrypted images from the ImageNet dataset were lower than those from the CIFAR-10 dataset.
\par 
Figure \ref{fig:4step_car} shows the examples of assembled images from the CIFAR-10 dataset, where Fig.\ref{fig:4step_car}\subref{fig:car_orig} is the original one.
It was confirmed that the scores of assembled images are low as the encypted images have a larger number of blocks.
As shown in Fig.\ref{fig:4step_car}\subref{fig:224_8_car}, although three step encryption and a smallest block size $B_{x}=B_{y}=8$ was used for the block-based image encryption, the largest part of the encrypted image was assembled due to the smaller number of blocks.
On the other hand, as shown in Fig.\ref{fig:4step_car}\subref{fig:384_8_car}, the use of a large number of blocks as $n=2304$ and a smallest block size $B_{x}=B_{y}=8$ enhances both invisibility and security against the jigsaw puzzle solver attack.
Examples of assembled images from the ImageNet dataset are shown in Fig. \ref{fig:4step_pen}, where Fig.\ref{fig:4step_pen}\subref{fig:pen_orig} is the original one.
As illustrated in Fig.\ref{fig:4step_pen}, the score of assembled image from the ImageNet dataset was far lower than that of the CIFAR-10 dataset.

\section{Conclusion}
In this paper, we evaluated the security of block-based image encryption for the vision transformer by using the jigsaw puzzle solver.
Experimental results showed that the use of three step encryption, and a smallest block size $B_{x}=B_{y}=8$ enhances robustness against the jigsaw puzzle solver attack.
Furthermore, it was confirmed that assembling encrypted image with 384$\times$384 pixels is much more difficult than 224$\times$224 pixels owing to the larger number of blocks.
\bibliographystyle{IEEETran}
\bibliography{refs}
\end{document}